# Challenge Theory: The Structure and Measurement of Risky Binary Choice Behavior

Samuel Shye[1] and Ido Haber[2,3]

### ABSTRACT

*Challenge Theory* (Shye & Haber 2015; 2020) has demonstrated that a newly devised *challenge index* (CI) attributable to every binary choice problem predicts the popularity of the *bold* option, the one of lower probability to gain a higher monetary outcome (in a gain problem); and the one of higher probability to lose a lower monetary outcome (in a loss problem). In this paper we show how Facet Theory structures the choice-behavior concept-space and yields rationalized measurements of gambling behavior. The data of this study consist of responses obtained from 126 student, specifying their preferences in 44 risky decision problems. A Faceted Smallest Space Analysis (SSA) of the 44 problems confirmed the hypothesis that the space of binary risky choice problems is partitionable by two binary axial facets: (a) Type of Problem (gain vs. loss); and (b) CI (Low vs. High). Four *composite variables,* representing the validated constructs: Gain, Loss, High-CI and Low-CI, were processed using Multiple Scaling by Partial Order Scalogram Analysis with base Coordinates (POSAC), leading to a meaningful and intuitively appealing interpretation of two necessary and sufficient gambling-behavior measurement scales.

### 1. Introduction

The Challenge Theory (CT) for decision under risk is a dual system stochastic model for binary choice behavior, based on the assumption that in decision under risk, two cognitive processing systems, the *automatic system* and the *analytic system* operate sequentially (Shye & Haber 2015; 2020). The automatic system reacts rapidly, providing the initial, *default* response which according to CT is based on the probabilities alone (initially disregarding the amounts of gain or of losses). Hence the default choice would be preference for the option to gain a smaller amount with a higher probability (in gains problems); and preference for the option to incur a higher loss with a lower probability (in loss problems). Then the analytic system enters into play examining the magnitude of the challenge involved in abandoning the default option and risking the alternative, the *bold*, option. The Challenge Index (CI) for every binary (monetary) choice problem was argued to be computable for both, gain and loss problems, thus:

$$CI[(x_0, p_0), (x_1, p_1)] = \frac{f_0|x_0|}{f_1|x_1|}(w_0(p_0) - w_1(p_1))$$

---

[1] The Van Leer Jerusalem Institute and the Hebrew University of Jerusalem.
[2] The Technion – Israel Institute of Technology.
[3] Research was partly supported by the National Mentoring Program, Szold Institute, Jerusalem.

Where for gain problems, in gamble $[(x_0, p_0), (x_1, p_1)]$, (with $x_1 > x_0 > 0$ and $0 < p_1 < p_0$), option $(x_0, p_0)$ is the default option and $(x_1, p_1)$ is the bold option. And where for loss problems, in gamble $[(x_0, p_0), (x_1, p_1)]$, (with $x_1 < x_0 < 0$ and $0 < p_1 < p_0$), option $(x_1, p_1)$ is the default option and $(x_0, p_0)$ is the bold option. And where $f_0$, $f_1$, $w_0$, $w_1$ are non-decreasing functions of their respective arguments, and $w_0$, $w_1$ are such that $w_0(p_0) - w_1(p_1) > 0$.

In Section 2 below, Facet Theory and its relevance to decision making research are briefly explained. Using Faceted Smallest Space Analysis (FSSA), Section 3 validates a set of gambling-behavior theoretical constructs. Section 4 describes the creation of composite variables for these constructs, taken to represent basic parameters of gambling behavior. The composite variables are analyzed by Partial Order Scalogram Analysis with base Coordinates (POSAC) to produce two scales that are necessary and sufficient for assessing gambling behavior.

## 2. Facet Theory for Decision-Making Research

As a meta-theory for multivariate behavioral research, Facet Theory (Guttman 1957; Shye, 1978; 1999; Shye & Elizur, 1995) aims to improve the stability (replicability) of scientific results by acknowledging that observed variables in behavioral research typically form but a *sample* from an infinite, or a very large number, of variables that make up the intended domain of investigation (the *content universe*). Consequently, Facet Theory proposes techniques for *sampling* variables (from the entire content universe) and making *inferences* (from the sampled variables to the entire content universe) as follows:

(a) *Sampling variables from the content universe.* This is done with the aid of a *mapping sentence*, a function whose domain consists of the respondents and of the stimuli as arguments, and whose image consists of the cartesian product of the ranges of responses to the stimuli, where each response-range is similarly ordered by a meaning (a concept) common to all stimuli. When stimuli are classified a priori by one or more content criteria, the mapping sentence facilitates stratified sampling of content universe[4].

A classification of the stimuli by their content is called *content facet*; and the response-range of a stimulus (classifying respondents by their response to that stimulus) is called a *range facet*.

(b) *Making inferences from the sample of variables to the entire content universe.* Such inferences require a specification of the kind of research-outcomes with respect to which inferences are to be made. Facet Theory posits that scientifically stable (replicable) outcomes would result from regional

---

[4] Aware of the problem-sampling issue, Erev et al. 2010 employ a random sampling procedure on their universe of choice problems, devising an algorithm for the random selection of the problem parameters (prizes and probabilities). While random sampling aims to cover the target problem-universe with respect to numerical problem-parameters, stratified sampling, employed in this study, aims to ensure adequate representation of problems with respect to each of several problem-classes (essentially qualitative criteria), where the specification of the classes is motivated by the theory being examined. The question of choosing the appropriate method for sampling variables in a particular study merits further attention.

hypotheses; hypotheses that specify a correspondence between (content- or range-) facets on the one hand, and partitionings into regions of certain geometric spaces, on the other hand[5] (see e.g., Shye 1978). Of the many spaces that have been proposed, two have proved especially fruitful:

*Faceted-SSA (Faceted Smallest Space Analysis;* Guttman, 1968; Shye & Elizur, 1994; Shye, 2014a). In the geometric space produced by this procedure, the objects of investigation are mapped as points, subject to the condition $r_{ij} > r_{kl} \Rightarrow d_{ij} \leq d_{kl}$, where $r_{ij}$ is a measure of similarity between object $i$ and $j$ (often the correlation coefficients between variables, if the objects are variables); and $d_{ij}$ is the distance between their points in space. The investigated universe, defined as the totality of possible objects of the type investigated, is represented by a topological manifold within the geometric space. Typically, the objects are variables, and in the present application, the variables are a sample of non-mixed monetary binary risky choice problems presented to respondents; the similarity measure is the pairwise correlation coefficient between variables; and the content universe is consists of all possible problems of the type investigated, with their response-ranges ordered each by the degree of *boldness* of the response. A regional hypothesis would then be, that a simple partition of the manifold can be found, such that each of its regions would include the variables of just one of the classes specified by the content facet. In this study the facets are a classification of choice problems by their *type* (gain vs. loss) and, independently, a classification of the problems by the *Challenge Index* (high vs. low). See Section 3 below.

*POSAC (Partial Order Scalogram Analysis by base Coordinates;* Shye, 1976; 1985; 2014b). This procedure is based on the analysis of investigated objects (often people) with respect to a partial order relationship that exists between them. Thus, it is assumed every pair of objects, $p_i, p_j$, is either *comparable* (designated by $p_i S p_j$) with one of them greater or equal ($p_i \geq p_j$) to the other; or they are *incomparable* (designated by $p_i \$ p_j$). POSAC aims to represent objects $p_i$ as points $x_1(p_i)\ldots x_m(p_i)$ in the smallest $m$-coordinates space, $X$, whose coordinates preserve observed partial order relations (comparability and incomparability). That is, $p_i \$ p_j$ iff for some two coordinates $x_s, x_t$

$$x_s(p_i) > x_s(p_j) \text{ but } x_t(p_i) < x_t(p_j).$$

A correspondence is then sought between each facet (typically, range facet) and a partition of the POSAC space into allowable regions, those that are separated by non-increasing hyperplanes (non-increasing lines, in the 2-dimensional case). See Sections 4 below.

---

[5] That is, where a one-to-one correspondence is specified between facet-elements (classes) and disjoint regions in space.

## 3. The Gambling Behavior Space: Faceted SSA

*3.1 Data*

The data of this study consist of responses obtained from 126 student, specifying their preferences in 44 risky decision problems, 22 gain problems and 22 loss problems, sampled from the universe of observations presented by the following mapping sentence:

**A Mapping Sentence for Risky Choice Behavior**

The extent to which respondent (x), faced with a binary risky $\begin{Bmatrix} Gain \\ Loss \end{Bmatrix}$ - **type** problem of $\begin{Bmatrix} High \\ Low \end{Bmatrix}$ **level** of challenge index (CI), responds with a bold gambling behavior $\rightarrow \begin{Bmatrix} High\ (chooses\ bold\ option) \\ Low\ (chooses\ default\ option) \end{Bmatrix}^n$ where *bold gambling behavior* is defined, for gain problems, as choosing the option of lower probability to make a higher gain (rather than the alternative option of higher probability to make a lower gain); and is defined for loss problems as choosing the option of higher probability to incur a lower loss (rather than the alternative option of lower probability to incur a higher loss). Note that, as required, variables defined in the mapping sentence all have a Common Meaning Range (CMR), namely that of *boldness*: The choice of the bold option indicates a gambling behavior that is high on boldness, and the choice of the default option indicates a gambling behavior that is low on boldness.

*3.2 Testing the Regional Hypotheses: The Structure of Gambling Behavior*

A two-dimensional Faceted SSA of the 44 problems (with the coefficient of weak monotonicity between reported boldness scores as similarity measure between problem-pairs) was performed using FSSAWIN (Figure 1).

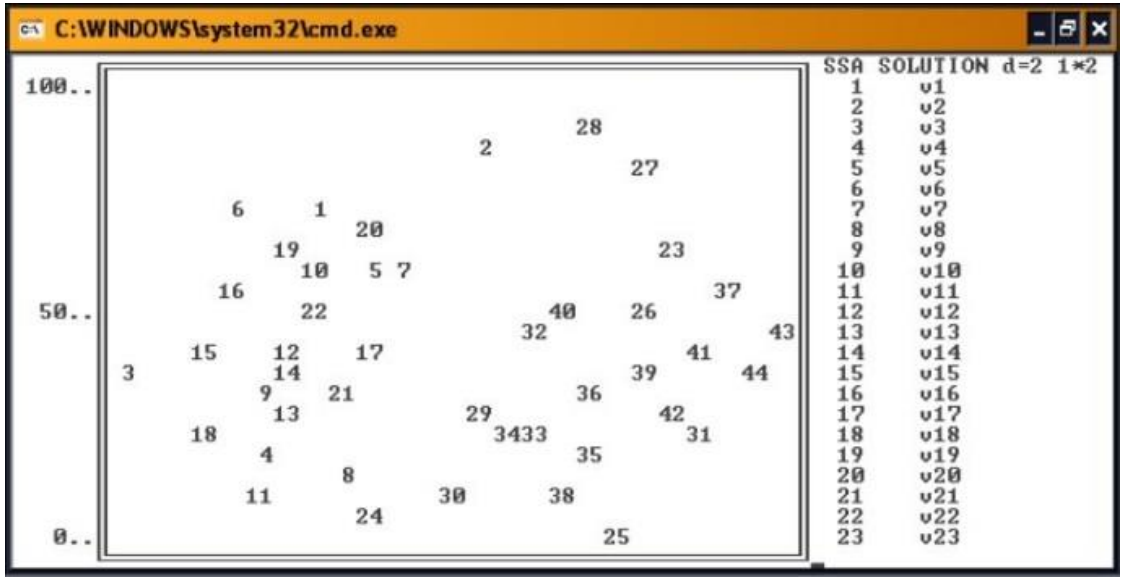

**Figure 1. Bold Gambling Behavior Space
Smallest Space Analysis of 44 Gambling Problems**
SSA Screen Diagram produced by FSSAWIN Program

Next, the regional hypotheses suggested by the problem-**type** and the problem-CI-**level** facets, were tested by Faceted SSA (FSSAWIN), a computerized partitioning procedure Shye (1991a, 1991b. See also Borg & Shye, 1995). Results confirm the two hypotheses: the problem-type facet is fully validated (Separation Index, SI=1.00); and the CI-level facet is well supported (SI=0.76). See Figure 2.

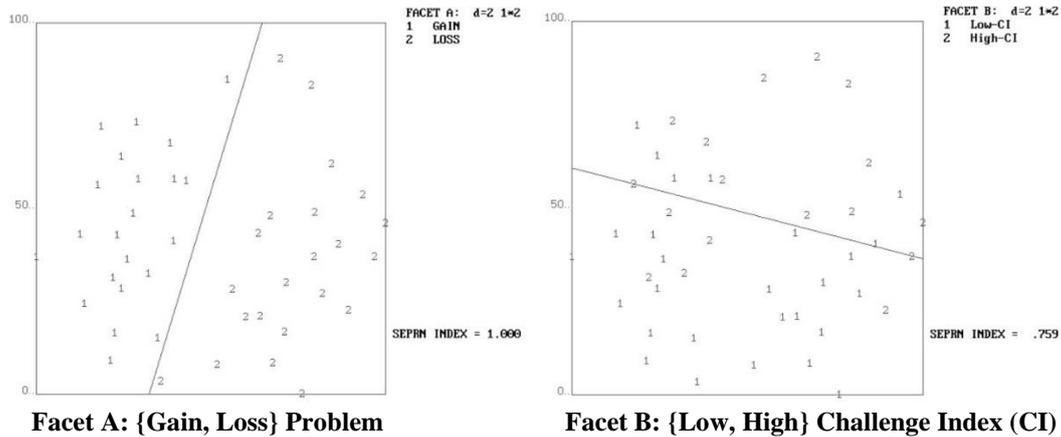

**Facet A: {Gain, Loss} Problem**     **Facet B: {Low, High} Challenge Index (CI)**
**Figure 2. Bold Gambling Behavior Space
Partitions by Facet A: {Gain, Loss} Problem and by Facet B: {High, Low} CI**
Partitioned Item Screen Diagrams produced by FSSAWIN Program

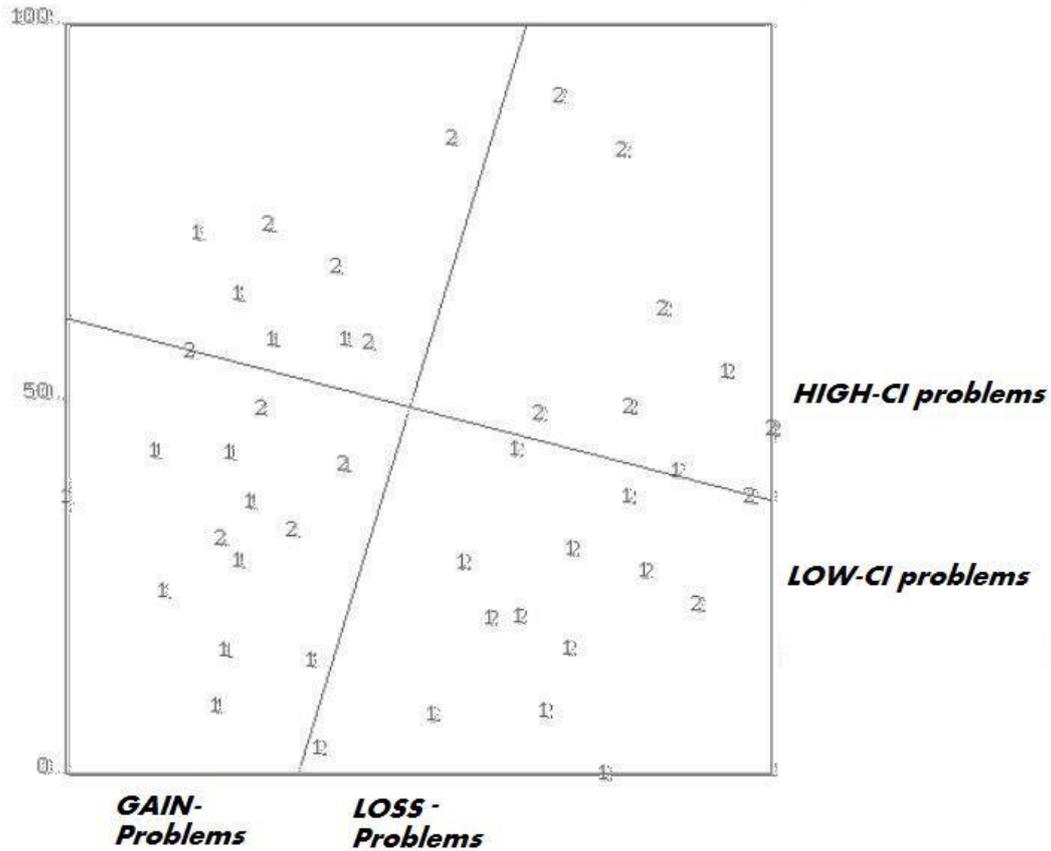

**Figure 3. The Structure of Bold Gambling Behavior Space
Partitions by both Facets, A: {Gain, Loss} and B: {High, Low} CI**
Superposition of the two Partitioned Item diagrams (by *Paint XP* software)

## 4. Multiple Scaling by POSAC of Gambling Behavior

*4.1 Constructs of gambling behavior*

The purpose of Multiple Scaling is better served if the variables analyzed by POSAC/LSA have been validated as representing theoretical constructs. Four theoretical constructs have been satisfactorily validated above by Faceted SSA. These constructs are:

*Preference of Bold Option in Gain Problems* (left-hand side in the map of the gambling behavior space; see Figure 4);
*Preference of Bold Option in Loss Problems,* (right-hand side in the map of the gambling behavior space; see Figure 4);
*Preference of Bold Option in Low-CI Problems* (top of the map of the gambling behavior space; see Figure 5).
*Preference of Bold Option in High-CI Problems,* (top of the map of the gambling behavior space; see Figure 5);

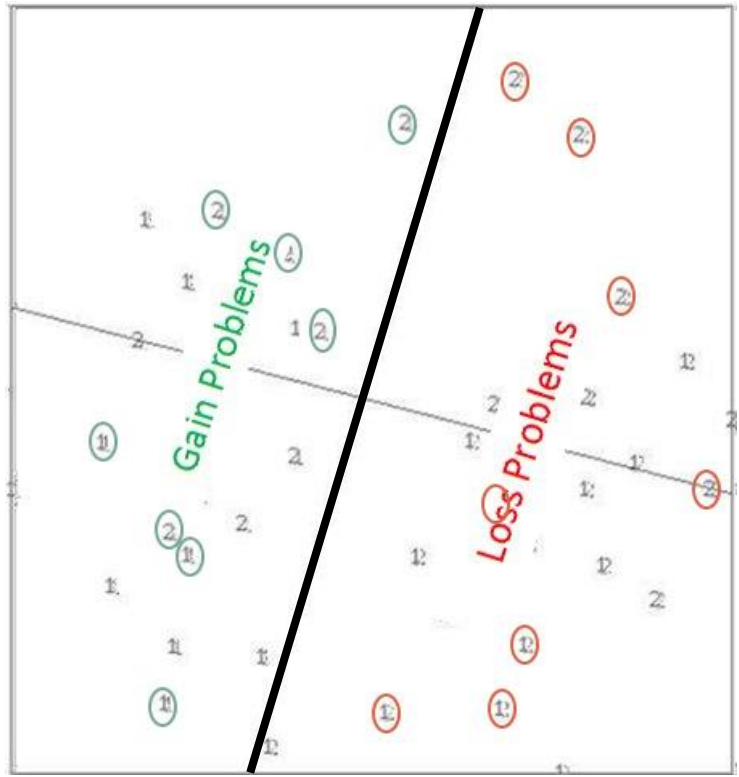

**Figure 4. Location in Bold Gambling Behavior Space of Variables Selected to Represent Gain Problem Region and Loss Problem Region**

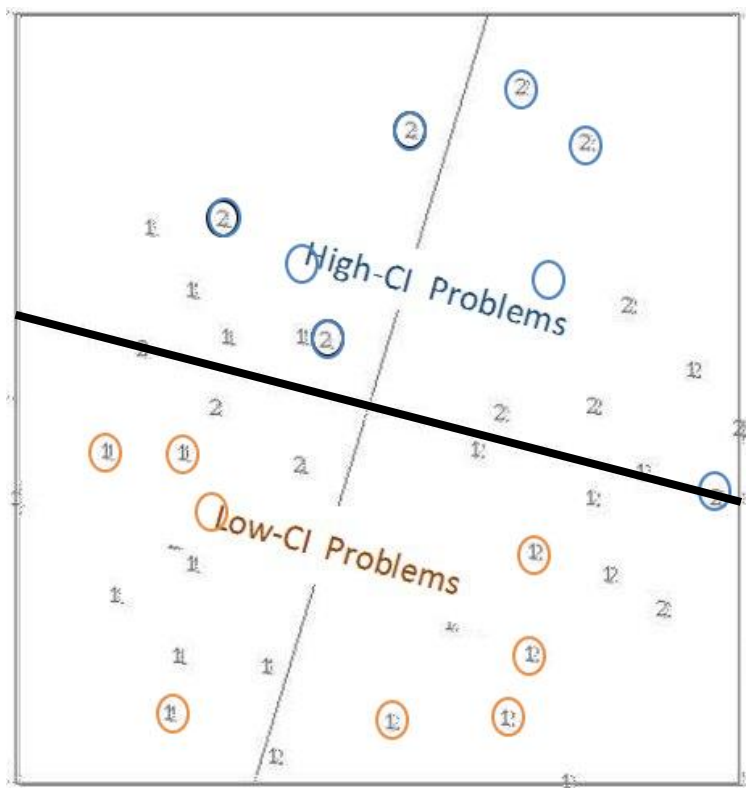

**Figure 5. Location in Bold Gambling Behavior Space of Variables Selected to Represent High-CI Problem Region and Low-CI Problem Region**

For each of these four constructs, taken to constitute a basic notion of gambling behavior, several representative variables were selected thus: Inspecting the region in the Bold Gambling-Behavior Space corresponding to the construct in question, those variables were identified which definitionally belong to that region; that is, variables that are not deviants. From among these variables, a set of variables was selected that was well spread over that region. The variables selected to represent a given construct were summed and dichotomized to form a *composite variable* for that construct, with 2 representing a bold choice and 1 representing a cautious (the default) choice, in each of the composite variables. Thus, four composite variables were created for respondents in the sample:

*Gain Composite Variable*. Representing the notion of *bold choice in gain problem*s;

*Loss Composite Variable*. Representing the notion of *bold choice in loss problems*;

*Lo-CI Composite Variable*. Representing the notion of *bold choice in low CI problems*.

*Hi-CI Composite Variable.* Representing the notion of *bold choice in high CI problems*;

*4.2 Multiple scaling by POSAC*

Every respondent in our sample now receives a profile of four dichotomous scores that represent his or her gambling behavior in terms of the validated constructs. These four-variable profiles, in turn, were processed by POSAC/LSA in order to discover whether a more parsimonious scaling may assess gambling behavior. A POSAC map of the 12 empirically obtained profiles, shown in Figure 6, displays a satisfactory partial-order representation of the 4-variable profiles in a 2-coordinate space (Correct Representation Coefficient= 0.97).

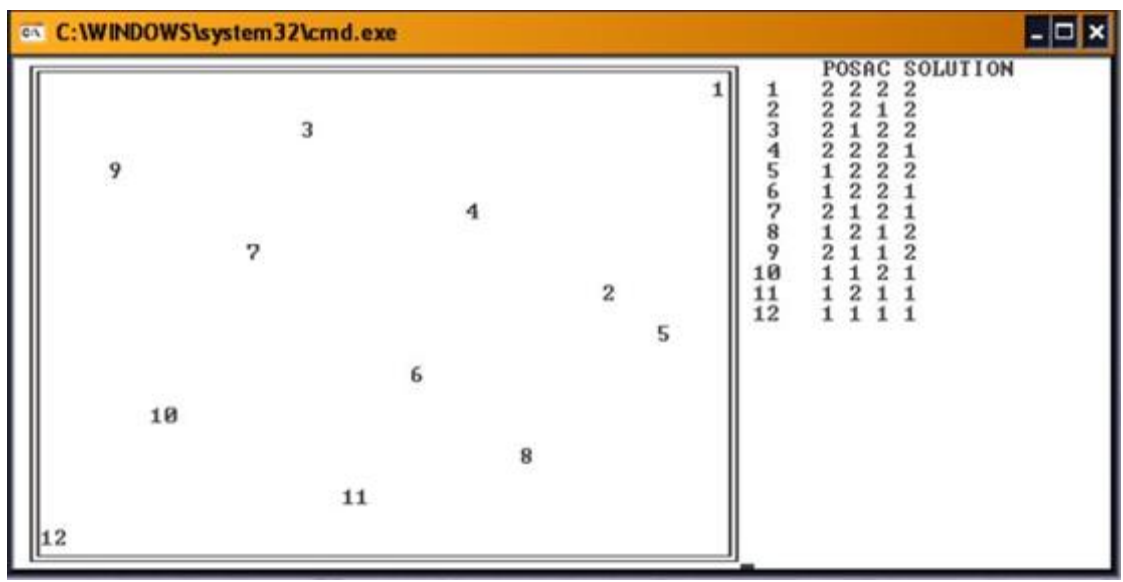

**Figure 6. POSAC Solution of the 4-Composite Variables Data**
Screen Diagram produced by POSAC/LSA Computer Program

Thus, the two coordinates constitute two scales whose scores are sufficient for reproducing the profiles for every subject in our sample. (Evidently, they are also necessary, since a single scale – a Guttman scale – will not do.) The scaling operation, however, requires also the *interpretation* of the two coordinate scales. That is, to reveal the substantive meaning of these two fundamental variables. This is achieved by investigating the *item diagrams* that POSAC produces. An item diagram, or a diagram of a given item (a variables processed) is a reproduction of the POSAC map, where, each profile-ID is replaced by the value of the item in that profile. For dichotomous items, partition lines are then sought that separate the profiles that are high (=2) in that item (above and to the right of the partition line) from those that are low (=1) in it (below and to the left of the partition line). It has been mathematically proven (Shye, 1976; 1985) that in two-dimensional POSAC the shape of those partition lines must be that of a decreasing (or non-increasing) step-curve; and that that there are exactly two items (variables) the shapes of whose partition lines are straight lines – one variable having a vertical partition line and the other having a horizontal partition line.

POSAC/LSA computer program finds for each item (POSAC-processed variable) the best fitting partition lines of four shapes: the best straight partition line (no bends); the best partition line with one bend (one right-angle step); the best partition line with two bends (two steps); and the best partition line with three bends (three steps). Each of these solutions is accompanied by a measure of goodness of fit; actually, by the total deviations of misfitting profiles from their region (taking into account the frequencies of those profiles).

An item whose partition line is straight, is said to be a *polar* item (or to play *polar role*). A polar item may be X-polar, if its partition-line crosses the X-axis (i.e., if it is vertical); or it may be Y-polar, if its partition-line crosses the Y-axis (i.e., if it is horizontal);

An item whose partition line has one bend, is said to be an *attenuating* item if its partition line is L-shaped; and it is said to be an *accentuating* item if its partition-line is inverted L shaped. Further roles are assigned to items with two or more bends. (See Shye, 1976; 1985)

A table produced by the program summarizes, for every item (variable), the sum total deviations for each of the four model partition lines. This information facilitates the determination of the optimal assignment of *roles* to the items. The deviations table obtained in the present study is shown in Figure 7.

**Figure 7. Deviations Table Produced by POSAC/LSA is Consulted to Determine the Optimal Role of Variables (Items) Processed by the Program**

Inspecting the table, we first identify the two polar items, those that have the minimum deviations from the model partition line. The choice in this case is simple: Item 1, representing the notion of "boldness in gain problems" is a Y-Polar item with 0 deviations from a (horizontal) straight line. Item 2, representing the notion of "boldness in loss problems", with its relatively small deviations for the polar is assigned the X-Polar role. (True, as an accentuating item it would fit perfectly, with deviations. But recall that there must be two polar items, and no other item would fit better as a polar item).

Next, we turn to examine the role played by Item 3, which represents the notion of "boldness in Low-CI problems". It cannot play a polar role because the best polar items have already been identified; and there can be no more than these two. Examining this item's fit as an accentuating/attenuating item (having a 1-bend partition line) we find that, indeed, by allowing a bend in its partition line, deviations decrease considerably from 354 to 109. Moreover, allowing further bends (2 or 3, corresponding to a *promoting role* or a *modifying role*) does not serve to reduce deviations. Hence the optimal role for Item 3 is attenuating role. (The T in the 5$^{th}$ column indicates *this, attenuating* role, rather than the alternative accentuating role, which is also associated with a 1-bend partition line.) Finally, Item 4, representing the notion of "Boldness in High-CI problems" clearly plays an accentuation with 0 deviations. (That it is an accentuating rather than an attenuating item, is indicated by the C in the 5$^{th}$ column of the table.)

For the present data, POSAC/LSA program produced 16 partitioned item diagrams (four partition-models for each of the four items). Here we reproduce only the four item diagrams identified above as optimal, one for each of the four items. These are the diagrams that will enables us to deduce the Gambling Behavior Measurement Space. See Figure 8 (a)-(d).

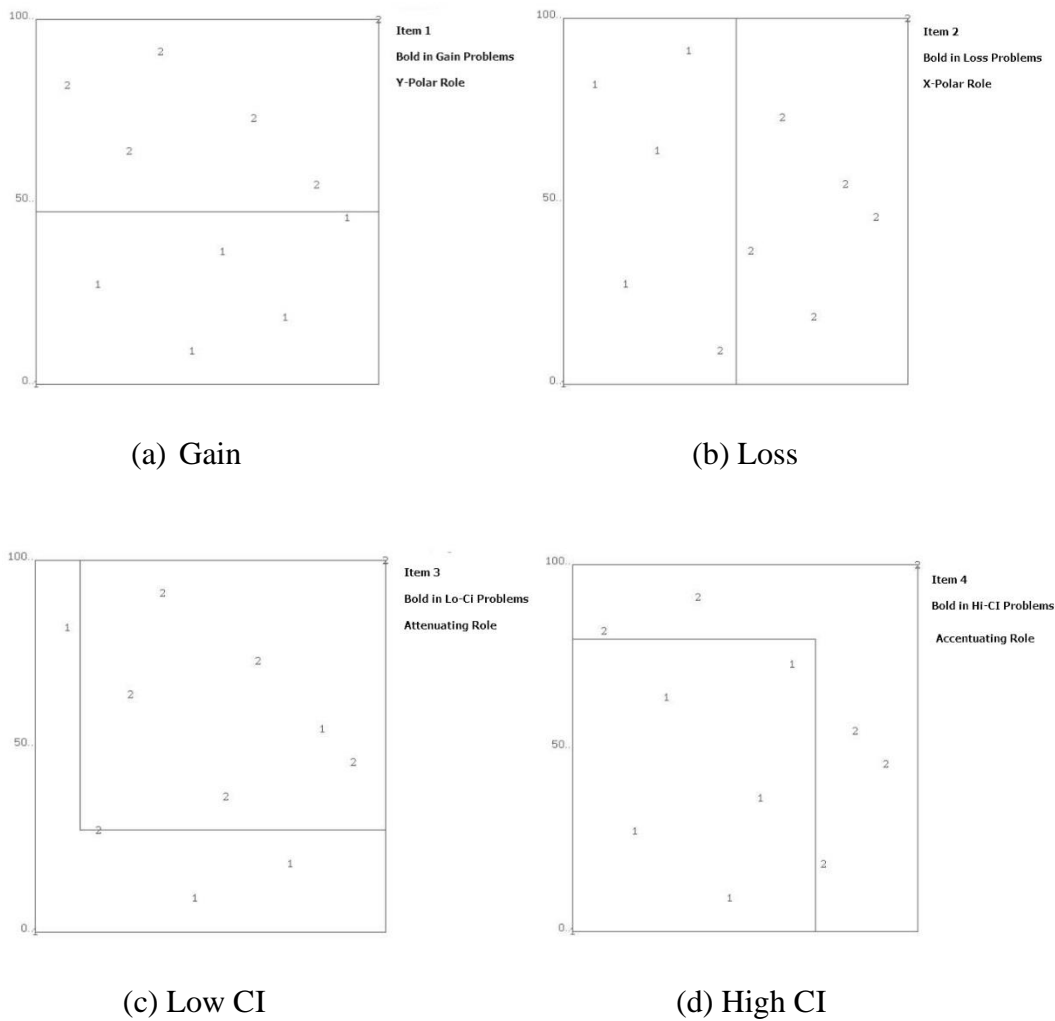

(a) Gain        (b) Loss

(c) Low CI        (d) High CI

**Figure 8. Optimal Partitioned Diagrams for the Four Items (Composite Variables: Boldness in Gain, Loss, High-CI, Low CI Problems)**

A superposition of the four item diagrams of Figure 8, constitutes the measurement space for gambling behavior (Figure 10). The two coordinates, X, Y, of this space are the two optimal – necessary and sufficient -- measurement scales for this behavior. Indeed, every profile-point in the POSAC space (based on the FSSA-validated constructs), is uniquely transformed into 2 scores, (x, y). The coordinate values (x, y) therefore reproduce the observed composite-variables profiles.

*4.3 Interpreting the coordinate-scales*

However, to complete this multiple-scaling procedure, the coordinate-scales must be interpreted; that is, their substantive contents must be determined. Being the fewest number of scales for the content universe studied, they are taken to constitute the fundamental variables of observed reality, the variables determine all empirically observed phenomena.

Using products of POSAC/LSA program, the interpretation of the scales is done by a piecemeal content-analysis of each of their intervals. Thus, the intervals in each coordinate must first be identified.

The interpretation procedure starts by noting the two polar variables, Gain and Loss. Starting, say, with the Gain Composite Variable, we find that low values in the Y-coordinate, those below the horizontal partition line, are associated with cautious gambling behavior (i.e., a relative tendency to prefer default options) in gain problems; while high values in the Y-coordinate, those above the horizontal partition line, are associated with bold choices (i.e., a tendency to prefer bold options) in gain problems. Similarly, the X-coordinate differentiates between cautious and bold gambling behavior in loss problems. The two polar variables enable a preliminary, rather gross, measurement of gambling behavior. See Figure 10. This 2-dimensional measurement space underscores the need for two rather than one scale: gambling behavior in loss problems differs in an essential way from that of gain problems. A unified, Guttman scale based on the notion of boldness alone will not do.

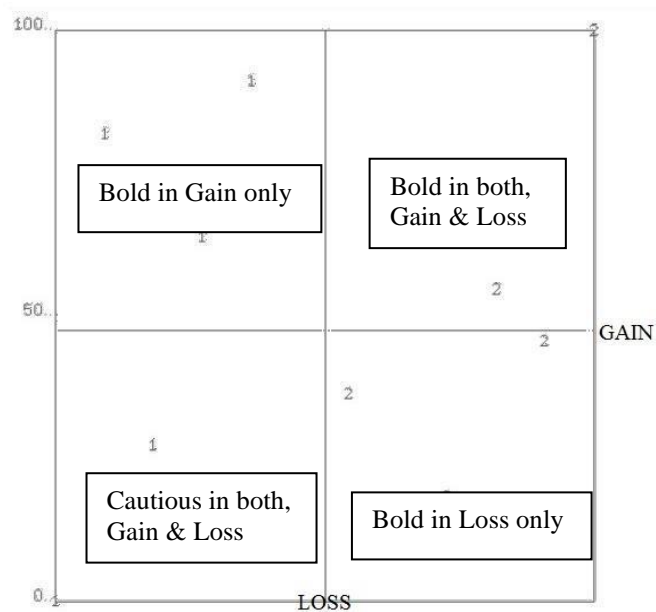

**Figure 9. Partitioning the Measurement Space by the Two Polar Items is the First Step in Interpreting the Gambling-Behavior Coordinate-Scales**

Next, we further partition the measurement space by the (L-shaped) partition line of Low-CI, and by the (Inverted-L-shaped) Partition line of High-CI, to obtain Figure 10. We note that the bend in the L-shaped partition line of Low-CI problems marks a point on the X-Coordinate and a point on the Y-Coordinate (see dotted lines). The result is a division of the X-Coordinate into four intervals (marked *1,2,3,4*) and similarly, a division of the Y-Coordinate into four intervals (marked *1,2,3,4*).

Drawing on the content-definitions of the four items that partition the space, we now interpret each of the two coordinates as scales of gambling behavior. The interpretation is essentially a semantic compound of the logical significance of the four intervals that make up the coordinate-scale.

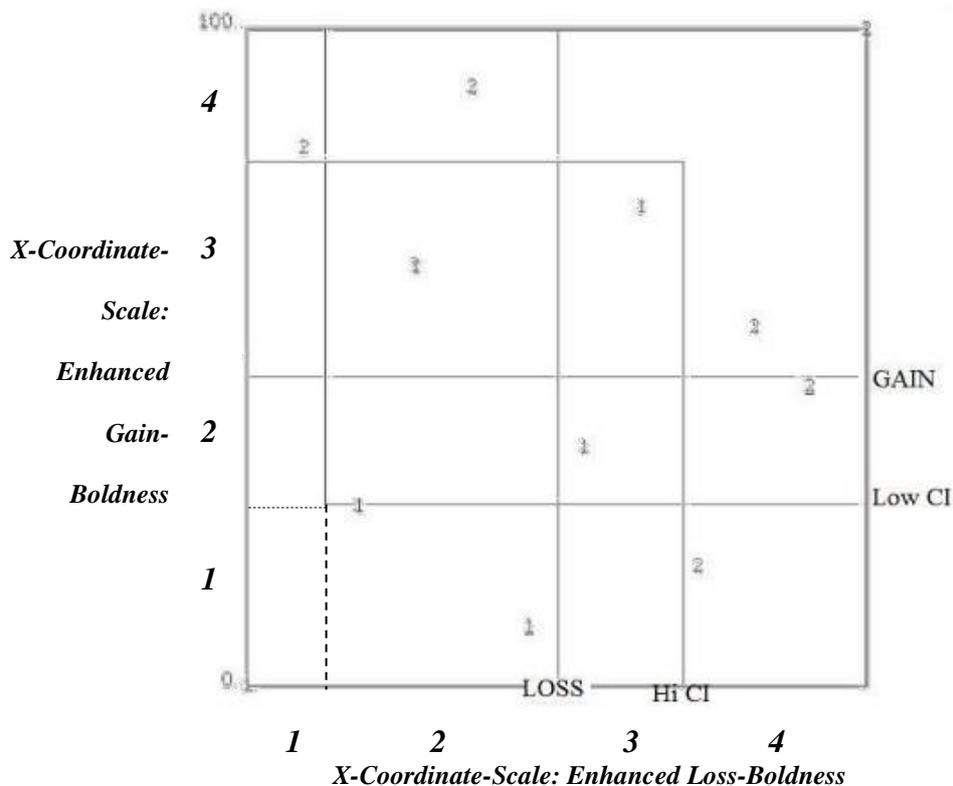

**Figure 10. Superposition of the Four Partitioned Diagrams of Figure 8 Determine Four Meaningful Intervals on the X-Coordinate Scale and Four Intervals on the Y-Coordinate Scale**

*4.3.1 Interpreting the X-Coordinate-Scale*

Intervals *1* and *2* represent cautious behavior in loss problems; for they characterize profiles that are to the left of (below) the vertical partition line of the "boldness in loss problems" construct. However, interval *1* represents, in addition, cautious behavior in Low-CI problems, while interval *2* represents bold behavior in Low-CI problems, thus alleviating the cautious behavior in loss problems (but not so much as to cross the threshold marked by the loss partition line). People (profiles) scoring *2* on the X-coordinate-scale are therefore bolder in loss problems than people with score *1* in that scale.

Intervals *3* and *4* represent bold behavior in loss problems; for they characterize profiles that are to the right (above) the vertical partition line of the "boldness in loss problems" construct. However, interval *3* represents, in addition, cautious behavior in High-CI problems, while interval *4* represents bold behavior in High-CI problems, thus strengthening the bold behavior in loss problems. People (profiles) scoring *4* on the X-coordinate-scale are therefore bolder in loss problems than people with score *3* in that scale.

We conclude that the X-coordinate-scale, with its four meaningful intervals, embodies a new concept of *enhanced loss-boldness-behavior.* This is an underlying fundamental variable that, together with Y-coordinate scale, enables preservation of observed order relations, including incomparability, among observed profiles.

*4.3.2 Interpreting the Y-Coordinate-Scale*

The interpretation of the Y-Coordinate-Scale follows the same logical steps as that of the X-Coordinate-Scale:

Intervals *1* and *2* represent cautious behavior in gain problems; for they characterize profiles that are below the horizontal partition line of the "boldness in gain problems" construct. However, interval *1* represents, in addition, cautious behavior in Low-CI problems, while interval *2* represents bold behavior in Low-CI problems, thus alleviating the cautious behavior in gain problems (but not so much as to cross the threshold marked by the gain partition line). People (profiles) scoring *2* on the Y-coordinate-scale are therefore bolder in gain problems than people with score *1* on that scale.

Intervals *3* and *4* represent bold behavior in gain problems; for they characterize profiles that are above the horizontal partition line of the "boldness in gain problems" construct. However, interval *3* represents, in addition, cautious behavior in High-CI problems, while interval *4* represents bold behavior in High-CI problems, thus strengthening the bold behavior in gain problems. People (profiles) scoring *4* on the Y-coordinate-scale are therefore bolder in gain problems than people with score *3* in that scale.

We conclude that the Y-coordinate-scale, with its four meaningful intervals, embodies a new concept of *enhanced gain-boldness-behavior.* This is another underlying fundamental variable that, together with X-coordinate scale, enables preservation of observed order relations, including incomparability, among observed profiles.

**5. Discussion**

Order relations, comparability as well as incomparability, among observed people's profiles constitute the essence of "measurement". The preservation of these relationships in the two-dimensional POSAC space means that a parsimonious gambling behavior measurement technique has been attained; and that the two scales, as fundamental variables, capture the essential factors that determine observed gambling behavior.

In many domains of research and applications, artificial intelligence (AI) procedures have been employed to process and draw conclusions concerning *numerical* or *pictorial* objects. The procedure described above, of inferring the semantic significance of the two scales, X and Y, may well be considered a novel AI procedure for processing and drawing conclusions concerning *semantic* objects.